\documentstyle[prd,aps,epsfig,floats]{revtex}
\begin{document}
\draft
\wideabs{
\title{Residual Entropy of Ordinary Ice from Multicanonical Simulations}

\author{Bernd A. Berg$^{\rm\,a,b,c}$, Chizuru Muguruma$^{\rm\,d}$, and 
        Yuko Okamoto$^{\rm\,c}$}

\address{ 
$^{\rm \,a)}$ Department of Physics, Florida State University,
  Tallahassee, FL 32306-4350, USA\\
$^{\rm \,b)}$ School of Computational Science, Florida State 
  University, Tallahassee, FL 32306-4120, USA\\
$^{\rm \,c)}$ Department of Physics, Nagoya University,
  Nagoya, Aichi 464-8602, Japan\\
$^{\rm \,d)}$ Faculty of Liberal Arts, Chukyo University,
  Toyota, Aichi 470-0393, Japan\\
} 
% (E-mail: berg@scs.fsu.edu)\\ 

\date{November 13, 2006} 
% \date{\today }

\maketitle
\begin{abstract}
We introduce two simple models with nearest neighbor interactions on 
3D hexagonal lattices. Each model allows one to calculate the residual
entropy of ice~I (ordinary ice) by means of multicanonical simulations.
This gives the correction to the residual entropy derived by Linus 
Pauling in 1935. Our estimate is found to be within less than 0.1\% of 
an analytical approximation by Nagle which is an improvement of Pauling's
result. We pose it as a challenge to experimentalists to improve on the 
accuracy of a 1936 measurement by Giauque and Stout by about one order 
of magnitude, which would allow one to identify corrections to Pauling's 
value unambiguously. It is straightforward to transfer our methods to 
other crystal systems. 
\end{abstract}
}
\narrowtext
% \section{Introduction}

A thorough understanding of the properties of water has a long history
and is of central importance for life sciences. After the discovery of
the hydrogen bond \cite{LaRo20} it was recognized that the unusual
properties of water and ice owe their existence to a combination of 
strong directional polar interactions and a network of specifically
arranged hydrogen bonds \cite{BeFo33}. The liquid phase of water 
differs from simple fluids in that there is a large qualitative
remnant of ice structure in the form of local tetrahedral 
ordering \cite{St92}.
 
\begin{figure}[-t] \begin{center} % WATER/work_water/Ice1/ice1_lat.f
\epsfig{figure=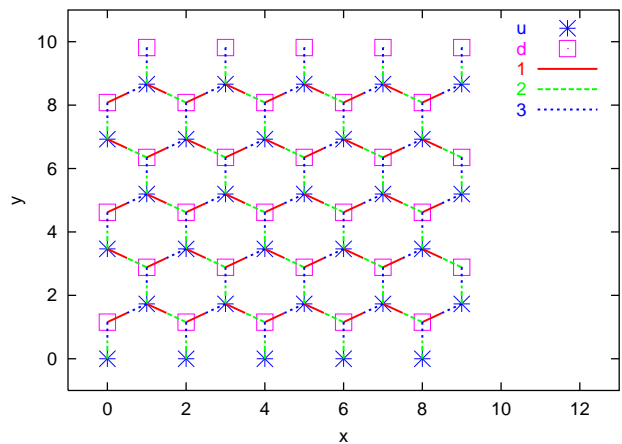,width=\columnwidth} \vspace{-1mm}
\caption{Lattice structure of one layer of ice~I. The up (u) sites 
are at $z=1/\sqrt{24}$ and the down (d) sites at $z=-1/\sqrt{24}$.
For each site three of its four pointers to nearest neighbor sites 
are shown. \label{fig_icepnt} }
\end{center} \vspace{-3mm} \end{figure}

In contrast to liquid water the properties of ice are relatively well
understood. Most of them have been interpreted in terms of crystal 
structures, the forces between its constituent molecules, and the 
energy levels of the molecules themselves \cite{EiKa69,PeWh99}. A 
two-dimensional projection of the hexagonal crystal structure of 
ordinary ice (ice~I) is depicted in Fig.~\ref{fig_icepnt} (other 
forms of ice occur in particular at high pressures). Each oxygen atom 
is located at the center of a tetrahedron and straight lines (bonds) 
through the sites of the tetrahedron point towards four nearest-neighbor 
oxygen atoms. Hydrogen atoms are distributed according to the ice rules 
\cite{BeFo33,Pa35}: (A)~There is one hydrogen atom on each bond (then 
called hydrogen bond). (B)~There are two hydrogen atoms near each 
oxygen atom (these three atoms constitute a water molecule). 

In our figure distances are given in units of a lattice constant $a$, 
which is chosen to be the edge length of the tetrahedra (this is
not the conventional crystallographic definition). For each 
molecule shown one of the surface triangles of its tetrahedron is 
placed in the $xy$-plane. The molecules labeled by u~(up) are then 
at $z=1/\sqrt{24}$ above, and the molecules labeled by d~(down) at 
$z=-1/\sqrt{24}$ below the $xy$-plane, at the centers of their 
tetrahedra. In our computer simulations information about the 
molecules will be stored in arrays of length $N$, $N$ being the 
number of molecules. 

Essentially by experimental discovery, extrapolating low temperature 
calorimetric data (then available down to about $10^{\circ}$K) towards 
zero absolute temperature, it was found that ice has a residual 
entropy \cite{Gi33}:
\begin{equation} \label{S0}
  S_0 = k\,\ln (W) > 0
\end{equation}
where $W$ is the number of configurations for $N$ molecules. 
Subsequently Linus Pauling \cite{Pa35} derived estimates of 
$W=(W_1)^N$ by two approximate methods, obtaining 
\begin{equation} \label{W1Pauling}
  W_1^{\rm Pauling} =  3/2  
\end{equation}
in each case. $W=(W_1)^N$ is the number of Pauling configurations. 
Assuming that the H$_2$O molecules are essentially intact in ice, 
his arguments are:

\begin{enumerate}
\item A given molecule can orient itself in six ways satisfying
      ice rule~B. Choosing the orientations of all molecules at
      random, the chance that the adjacent molecules permit a
      given orientation is 1/4. The total number of configurations
      is thus $W=(6/4)^N$.
\item Ignoring condition B of the ice rules, Pauling allows $2^{2N}$ 
      configuration on the hydrogen bonds between adjacent oxygen 
      atoms: Each hydrogen nucleus is given the choice of two 
      positions, near to one of the  two oxygen atoms. At one
      oxygen atom there are now sixteen arrangements of the four 
      hydrogen nuclei. Of those ten are ruled out by ice rule~B. 
      This condition for each oxygen atom permits $6/16=3/8$ 
      of the configurations. Accordingly, the total number of 
      configurations becomes $W=2^{2N}\,(3/8)^N$.
\end{enumerate}

Equation~(\ref{W1Pauling}) converts to the residual entropy
\begin{equation} \label{S0Pauling}
  S_0^{\rm Pauling} = 0.80574\dots\ {\rm cal/deg/mole}
\end{equation}
where we have used $R=8.314472\,(15)\, [{\rm J/deg/mol}]$ for
the gas constant as of Aug.~20, 2006 given on the NIST website 
\cite{NIST} (relying on CODATA 2002 recommended values). This
in good agreement with the experimental estimate
\begin{equation} \label{S0experimental}
  S_0^{\rm experimental} = 0.82\, (5)\ {\rm cal/deg/mole}
\end{equation}
which was subsequently obtained by Giauque and Stout~\cite{Gi36}
using refined calorimetry (we give error bars with respect to the 
last digit(s) in parentheses).

Pauling's arguments omit correlations induced by closed loops when 
one requires fulfillment of the ice rules for all atoms, and it was 
shown by Onsager and Dupuis \cite{OnDu60} that $W_1=1.5$ is in fact 
a lower bound. Onsager's student Nagle used a series expansion method 
to derive the estimate \cite{Na65}
\begin{equation} \label{W1Nagle}
  W_1^{\rm Nagle} = 1.50685\, (15)\,,
\end{equation}
or
\begin{equation} \label{S0Nagle}
  S_0^{\rm Nagle} = 0.81480\, (20)\ {\rm cal/deg/mole}\,.
\end{equation}
Here, the error bar is not statistical but reflects higher order 
corrections of the expansion, which are not entirely under control.
The slight difference between (\ref{S0Nagle}) and the value in Nagle's
paper is likely due to improvements in the measurements of Avogadro's 
number~\cite{NIST}. The only independent theoretical value appears to 
be one for cubic ice, which is obtained by numerical integration 
of Monte Carlo data \cite{Is04} and in good agreement with Nagle 
\cite{Na65}. 

Despite Nagle's high precision estimate there has apparently been 
almost no improvement on the accuracy of the experimental value 
(\ref{S0experimental}). Some of the difficulties are addressed in 
a careful study by Haida et al. \cite{Ha74}. But their final estimate
remains (\ref{S0experimental}) with no reduction of the error bar. 
We noted that by treating the contributions in their table~3 as 
statistically independent quantities and using Gaussian error 
propagation (instead of adding up the individual error bars), the 
final error bar becomes reduced by almost a factor of two and
their value would then read $S_0=0.815\,(26)\,$cal/deg/mol. Still
Pauling's value is safely within one standard deviation. Modern
electronic equipment should allow for a much better precision. We
think that an experimental verification of the difference to
Pauling's estimate would be an outstanding confirmation of structures 
imposed by the ice rules.

In this paper we provide a novel high-precision numerical estimate 
of $S_0$ for ordinary ice. Our calculations are based on two simple 
statistical models, which reflect Pauling's arguments. Each model is 
defined on the hexagonal lattice structure of Fig.~\ref{fig_icepnt}. 

In the first model, called 6-state H$_2$O molecule model, we allow 
for six distinct orientations of each H$_2$O molecule and define its 
energy by
\begin{equation} \label{E1}
  E = - \sum_b h(b,s^1_b,s^2_b)\ .
\end{equation}
Here, the sum is over all bonds $b$ of the lattice and ($s^1_b$ and 
$s^2_b$ indicate the dependence on the states of the two H$_2$O 
molecules, which are connected by the bond)
\begin{equation} \label{hs}
  h(b,s^1_b,s^2_b) = \cases{1\ {\rm for\ a\ hydrogen\ bond},\cr 
                            0\ {\rm otherwise}. }
\end{equation}

In the second model, called 2-state H-bond model, we do not consider 
distinct orientations of the molecule, but allow two positions for each 
hydrogen nucleus on the bonds. The energy is defined by
\begin{equation} \label{E2}
  E = - \sum_s f(s,b^1_s,b^2_s,b^3_s,b^4_s)\,,
\end{equation}
where the sum is over all sites (oxygen atoms) of the lattice. The 
function $f$ is given by
\begin{eqnarray} \label{fs}
  &~&f(s,b^1_s,b^2_s,b^3_s,b^4_s) = \\ \nonumber &~&
  \cases{2\ {\rm for\ two\ hydrogen\ nuclei\ close\ to}\ s, \cr 
  1\ {\rm for\ one\ or\ three\ hydrogen\ nuclei\ close\ to}\ s,\cr 
  0\ {\rm for\ zero\ or\ four\ hydrogen\ nuclei\ close\ to}\ s. }
\end{eqnarray}

The groundstates of each model fulfill the ice rules. At $\beta=0$ the 
number of configurations is $6^N$ for the 6-state model and $2^{2N}$ 
for the 2-state model. Because the normalizations at $\beta=0$ are 
known, multicanonical (MUCA) simulations \cite{Be92a} allow us 
in either case to estimate accurately the number of groundstate 
configurations \cite{Be92b}. Superficially both systems resemble Potts 
models (see, e.g., \cite{BBook} for Potts model simulations), but their 
thermodynamic properties are entirely different. For instance, we 
do not find any sign of a disorder-order phase transition, which is 
for our purposes advantageous as the MUCA estimates for the groundstate 
entropy become particularly accurate. This  absence of a bulk transition
does not rule out long-range correlations between bonds of the ground
state configurations, which are imposed by the conservation of the
flow of hydrogen bonds at each molecule. In that sense the ground
state is a critical ensemble.

Using periodic boundary conditions (BCs), our simulations are based
on a lattice construction set up earlier by one of us~\cite{Be05}. 
Following closely the method outlined in chapter~3.1.1 of \cite{BBook} 
four index pointers from each molecule to the array positions of its 
nearest neighbor molecules are constructed. The order of pointers one 
to three is indicated in Fig.~\ref{fig_icepnt}. The fourth pointer is 
up the $z$-direction for the u molecules and down the $z$-direction 
for the d molecules. The lattice contains then $N=n_x\,n_y\,n_z$ 
molecules, where $n_x$, $n_y$, and $n_z$ are the number of sites
along the $x$, $y$, and $z$ axes, respectively. The periodic BCs 
restrict the allowed values of $n_x$, $n_y$, and $n_z$ to $n_x = 
1,\,2,\,3,\,\dots$, $n_y = 4,\,8, \,12,\,\dots$, and $n_z = 2,\,4,\,6,
\,\dots$~. Otherwise the geometry does not close properly. Using the 
inter-site distance $r_{OO}=2.764\,$\AA\ from Ref.~\cite{PeWh99}, 
the physical size of the box is obtained by putting the lattice 
constant $a$ to $a=2.257\,$\AA , and the physical dimensions of the 
box are calculated to be $B_x=2n_x\,a$, $B_y=(n_y\,\sqrt{3}/2)\,a$, 
$B_z=(n_z\,4/\sqrt{6})\,a$. In our choices of $n_x$, $n_y$, and 
$n_z$ values we aimed within reasonable limitations at symmetrically 
sized boxes.

\begin{table}[tb]
\caption{ Simulation data for $W_1$. \label{tab_data}} 
% Results in 
\centering
\begin{tabular}{|c|c|c|c|c|c|c|} 
 $N$&$n_x$&$n_y$&$n_z$& 6-state model & 2-state model& $Q$\\   \hline
 &&&&$\qquad W_1\qquad N_{\rm cyc}$&$\qquad W_1\qquad N_{\rm cyc}$&\\ 
\hline
 128&   4 &   8 &   4 &$1.52852\,(47)$ 1854&$1.52869\,(23)$ 7092&0.72\\ 
\hline
 360&   5 &  12 &   6 &$1.51522\,(49)$~~223&$1.51546\,(15)$ 1096&0.65\\
\hline
 576&   6 &  12 &   8 &$1.51264\,(18)$~~503&$1.51279\,(10)$ 1530&0.47\\
\hline
 896&   7 &  16 &   8 &$1.51075\,(16)$~~208&$1.51092\,(06)$ 2317&0.32\\
\hline
1600&   8 &  20 &  10 &$1.50939\,(09)$~~215&$1.50945\,(05)$~~619&0.56\\
\hline
\multicolumn{3}{|c}{$\infty$ (fit)}&  
                      &$1.50741\,(33)\qquad$&$1.50737\,(17)\qquad$& 0.91\\
\end{tabular} \end{table} % \vspace*{0.2cm}

Table~\ref{tab_data} compiles our MUCA $W_1$ estimates for the lattice 
sizes used. In each case a Wang-Landau recursion \cite{WL01} was used 
to estimate the MUCA parameters for which besides a certain number of 
cycling events \cite{BBook} a flatness of $H_{\min}/H_{\max}>0.5$ 
was considered sufficient for stopping the recursion and starting
the second part of the MUCA simulation with fixed weights ($H(E)$ is 
the energy histogram, $H_{\min}$ is the smallest and $H_{\max}$ the 
largest number of entries in the flattened energy range).

The statistics we used for measurements varied between $32\times 10^6$
sweeps for our smallest and $64\times 10^7$ sweeps for our largest 
lattice. Using two 2~GHz PCs the simulations take less than one week. 
The number of cycles, $N_{\rm cyc}$, completed between $\beta_{\min}=0$ 
and the groundstate are listed in the 6-state and 2-state model columns 
of table~\ref{tab_data}. 
As each of our simulations includes the $\beta_{\min}=0$ canonical 
ensemble, the (logarithmically coded) re-weighting procedure of 
chapter~5.1.5 of \cite{BBook} delivers estimates for $W_1$, which 
are compiled in the same columns. Each error bar relies on 32 jackknife 
bins. As expected, the values from both models are consistent as is 
demonstrated by $Q$ values of Gaussian difference tests (see, e.g., 
chapter~2.1.3 of \cite{BBook}) in the last column of the table. 
The 2-state H-bond model gives more accurate estimates than the
6-state H$_2$O molecule model, obviously by the reason that the cycling 
time, which is $\propto 1/N_{\rm cyc}$, is less for the former, because 
the energy range that needs to be covered is smaller.

\begin{figure}[-t] \begin{center} % WATER/work/Ice1ResEntropy/fit1N.plt
\epsfig{figure=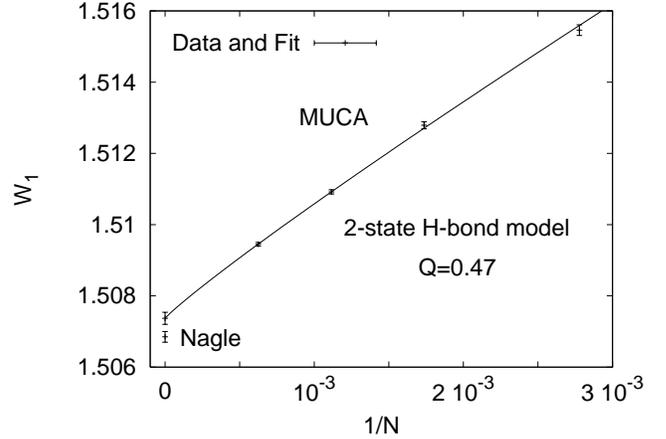,width=\columnwidth} \vspace{-1mm}
\caption{Fit for $W_1$. \label{fig_fit} }
\end{center} \vspace{-3mm} \end{figure}

In Fig.~\ref{fig_fit} a fit for the data of the 2-state H-bond model
to the form 
\begin{equation} \label{fit}
 W_1(x) = W_1(0) + a_1\,x^{\theta}\,,~~~x=1/N
\end{equation}
is shown. The $W_1=W_1(0)$ estimate from Fig.~\ref{fig_fit} is given in 
the the last row of the 2-state model column of table~\ref{tab_data}. 
The data point for the smallest lattice is included in the fit, but 
not shown in the figure where we like to focus on the large $N$ region. 
The goodness of fit (chapter~2.8 of \cite{BBook}) is $Q=0.47$ as given 
in the figure.  Similarly the estimate for the 6-state H$_2$O molecule 
model in the last row of the table is obtained with a goodness of fit 
$Q=0.78$. All $Q$ values (Gaussian difference tests and fits) are in 
the range one would expect for statistically consistent data. The 
$\theta$ values of the fits were also consistent and their combined 
value is $\theta = 0.923\,(23)$. That we have $\theta\ne 1$ reflects 
bond correlations in the groundstate ensemble.

Combining the two fit results weighted by their error bars leads
to our final estimate
\begin{equation} \label{W1MUCA}
  W_1^{\rm MUCA} = 1.50738\, (16)\,.
\end{equation}
This converts into
\begin{equation} \label{S0MUCA}
  S_0^{\rm MUCA} = 0.81550\, (21)\ {\rm cal/deg/mole}\,.
\end{equation}
for the residual entropy.

The difference between (\ref{W1MUCA}) and the estimate of Nagle 
(\ref{W1Nagle}) is 0.035\% of the estimated $W_1$ value (0.086\% of
$S_0$), which is much smaller than any foreseeable experimental error. 
However, within their own error bars the Gaussian difference test 
between the two estimates yields $Q=0.016$. As the error bar in 
(\ref{W1MUCA}) covers and no systematic errors due to finite size 
corrections from larger lattices, the small discrepancy with Nagle's 
result may well be explained this way. In view of the large error bar 
in the experimental estimate it appears somewhat academic to trace 
the ultimate reason.

As already (hesitatingly) pointed out by Pauling \cite{Pa35}, the real 
entropy at zero temperature is not expected to agree with the residual 
entropy extrapolated from low but non-zero temperatures. In real ice 
one expects a small splitting of the energy levels of the Pauling 
configuration, which are degenerate in both of our models. Once the
thermal fluctuations become small compared with these energy differences,
the entropy will become lower than the residual entropy calculated
here. Such an effect is observed in \cite{Ha74} by annealing ice~I at 
temperatures between $85^{\circ}$K and $110^{\circ}$K. Refined models 
are needed to gain computational insights. Crossing this temperature 
range sufficiently fast allows one still to extract the residual 
entropy, because the relaxation time has become so long that one 
does not have ordering of Pauling states during typical experimental 
observation times.

It is clear that our method carries rather easily over to other crystal 
structures for which one may want to calculate residual entropies. In 
particular structural defects and impurities can be included, although 
one may have to use more realistic energy functions and lattice sizes 
put limits on low densities. Simulations very similar to those performed 
here should enable accurate estimates of the residual entropies for 
other forms of ice and various geometrically frustrated systems 
\cite{Hi04} as well as for spin models in the class for which lower 
bounds on their residual entropies were derived in \cite{ChWu87}. For 
more involved systems our approach is to design simple models, which 
share the relevant groundstate symmetries with the system of interest. 
That could, for instance, have applications to the residual entropy 
of proteins by allowing for more realistic modeling than done in
Ref.~\cite{Li96}.

Finally, good modeling of water is of crucial importance for computational
progress in biophysics. Clusters of hydrogen bonds play a prominent
role in water at room temperature. Our method allows one to calculate 
the combinatorial factors $W_1^N$ with which small clusters ought to 
be calculated in phenomenological water models like those discussed 
in Ref.~\cite{St92}. Through a better understanding of hydrogen bond
clusters insights derived from the study of ordinary ice may well 
be of importance for improving on models \cite{Models}, which have 
primarily been constructed to reflect properties of water under room 
temperatures and pressures.

\acknowledgments
We like to thank John Nagle for e-mail comments on the first version
of this paper.
During most of this work Bernd Berg was supported by a fellowship of
the JSPS. Chizuru Muguruma and Yuko Okamoto were supported, in part, 
by the Ministry of Education, Culture, Sports, Science and Technology
(MEXT), Japan: Chizuru Muguruma by Grants-in-Aid for Young 
Scientists~(B), Grant No. 16740244 and Yuko Okamoto by Grants-in-Aid 
for the Next Generation Super Computing Project, Nanoscience Program 
and for Scientific Research in Priority Areas, Water and Biomolecules.

\clearpage
\end{document}